\documentclass[manuscript]{aastex}

\newcommand{\Hubble}{{\it Hubble Space Telescope}}
\newcommand{\HST}{{\it HST}}

\def\pmb#1{\setbox0=\hbox{#1}
  \kern-.02em\copy0\kern-\wd0
  \kern.01em\copy0\kern-\wd0
  \kern.01em\copy0\kern-\wd0
  \kern.01em\copy0\kern-\wd0
  \kern.01em\copy0\kern-\wd0
  \kern-.02em\raise.01em\box0 }

\def\ref#1#2{$^{#1}$}

\shorttitle{V838 Mon Light Echo}
\shortauthors{Bond et al.}

\begin{document}



\title{An energetic stellar outburst accompanied by circumstellar light
echoes}

\author{Howard E. Bond\altaffilmark{1},
Arne Henden\altaffilmark{2}, 
Zoltan G. Levay\altaffilmark{1}
Nino Panagia\altaffilmark{1,3},
William~B.~Sparks\altaffilmark{1}, 
Sumner Starrfield\altaffilmark{4}, 
R. Mark Wagner\altaffilmark{5}, 
R. L. M. Corradi\altaffilmark{6},
\& U.~Munari\altaffilmark{7}
}

\altaffiltext{1}
{Space Telescope Science Institute, 
3700 San Martin Dr.,
Baltimore, MD 21218, USA}

\altaffiltext{2} {Universities Space Research Association \& US Naval
Observatory, Flagstaff Station, P.O. Box 1149, Flagstaff, AZ 86002, USA }

\altaffiltext{3}
{On assignment from European Space Agency, Space Telescope Division}

\altaffiltext{4}
{Dept.\ of Physics \& Astronomy, Arizona State University,
Tempe, AZ 85287-1504, USA}

\altaffiltext{5}  
{Large Binocular Telescope Observatory, University of Arizona, 933 North Cherry
Ave., Tucson, AZ 85721, USA}

\altaffiltext{6} 
{Isaac Newton Group of Telescopes, Apartado de Correos 321,
38700 Santa Cruz de La Palma, Canarias, Spain}

\altaffiltext{7}  
{INAF-Osservatorio Astronomico di Padova, Sede di Asiago, 36012 Asiago (VI),
Italy}

\slugcomment{\bf To appear in Nature, March 27, 2003}

\clearpage

\begingroup \bf

Some classes of stars, including supernovae and novae, undergo explosive
outbursts that eject stellar material into space.  In 2002, the previously
unknown variable star V838 Monocerotis brightened suddenly by a factor of
\pmb{$\sim$$10^4$}.  Unlike a supernova or nova, V838~Mon did not explosively
eject its outer layers; rather, it simply expanded to become a cool supergiant
with a moderate-velocity stellar wind.   Superluminal light echoes were
discovered\ref{1,2}{(Henden et al. 2002; Munari et al. 2002)} as light from the
outburst propagated into surrounding, pre-existing circumstellar dust.  Here we
report high-resolution imaging and polarimetry of the light echoes, which allow
us to set direct geometric distance limits to the object.  At a distance of
$>$6~kpc, V838~Mon at its maximum brightness was temporarily the brightest star
in the Milky Way. The presence of the circumstellar dust implies that previous
eruptions have occurred, and spectra show it to be a binary system. When
combined with the high luminosity and unusual outburst behavior, these
characteristics indicate that V838~Mon represents a hitherto unknown type of
stellar outburst, for which we have no completely satisfactory physical
explanation.

\endgroup

Light echoes around an outbursting star appear after the eruption, the
illumination being delayed by the longer path length from the star to the
surrounding dust and then to the Earth, as compared with the light from the
outburst itself traveling directly to the Earth. In the case of an
instantaneous light pulse, the geometry of a light echo is
simple\ref{3-8}{(Kapteyn 1902; Couderc 1939; Russell, Dugan \& Stewart 1945;
Chevalier 1986; Schaefer 1988; Sparks 1994)}: the illuminated dust lies on the
paraboloid given by $z=x^2/2ct-ct/2$, where $x$ is the projected distance from
the star in the plane of the sky, $z$ is the distance from this plane along the
line of sight toward the Earth, $c$ is the speed of light, and $t$ is the time
since the outburst.  Light echoes are rare, having been observed only around a
few classical novae in our own Galaxy\ref{3,9,10}{(Schaefer 1988; Ritchey 1902;
Casalegno et al.\ 2000)}, and around several extragalactic
supernovae\ref{11}{(Sparks et al.\ 1999)}. With the exception of
SN~1987A\ref{12}{(Bond et al.\ 1990)}, these echoes arose from  interstellar
dust or gas in the line of sight, rather than from true circumstellar material.

V838 Mon erupted in early 2002 January\ref{13}{(Brown 2002)}. Fig.~1 shows the
light curve of the outburst in linear flux units, from ground-based
observations in three different wavelength bands: blue ($B$), visual ($V$), and
near-infrared ($I$)\null.  This light curve---with a weak precursor
brightening, a sharp, bright blue peak, followed by a decline and then two
further broad and much redder peaks---is unlike that of a supernova, nova, or
any other known type of variable star.   Spectra reported in numerous {\it IAU
Circulars\/} showed a very cool photosphere throughout the outburst. Toward the
end of the outburst, and during the decline, the star became extremely red.
Recent spectra\ref{20-22}{(Wagner \& Starrfield 2002; Munari \& Desidera 2002;
Geballe et al. 2002)}, taken after the decline, show that a hot B-type star is
also present, along with the cool outbursting object. 

We have obtained \Hubble\/ (\HST) observations of the light echoes surrounding
V838~Mon with the Advanced Camera for Surveys (ACS) at the four epochs listed
in Table~1.  Fig.~2 shows the ACS images. Only the $B$ filter was used at the
first epoch, but for the others we obtained three-color ($BVI$) frames, and
have used these to make renditions that approximate true color. The nebulosity
exhibits a wealth of subarcsecond cirrus-like structure, but is dominated
(especially at the first two epochs) by a remarkable series of nearly circular
arcs and rings, centered on the variable star. The exact linear size of the
illuminated dust envelope depends on the distance to the object, but based on
the distance limits given below, the visible material is confined to less than
2~pc from the star and is thus circumstellar, rather than being ambient
interstellar medium. The cavity centered on the star, e.g.\ in Figs.~2a and 2b,
along with several features that appear aligned toward the star, also establish
the circumstellar nature of the echoes.

As shown in Fig.~1, the outburst light curve had the simplest structure in
$B$---a bright, relatively short peak on February~6, and a second, weaker peak
about 32 days later---making the $B$-band light echoes most amenable to
geometric analysis. If we assume the simplest geometry, which is a series of
nested spherical dust shells centered on the star and intersecting the
expanding light paraboloids, then it can be shown that the apparent 
expansion rate of a ring is given by $dx/dt = (rc - c^2t) (2rct -
c^2t^2)^{-1/2}$, where $r$ is the radius of the ring measured from the star.
Thus the time behavior of these rings in angular units {\it provides a direct
geometric distance to the object}.  The pair of images taken only 20.8~days
apart, on April 30 and May 20, is the most suitable for such a measurement,
since the same rings can be identified unambiguously in both images.   The
expansion rates are plotted against the mean angular radii of each ring in
Fig.~3.  Overlaid on the plot are the expansion rates calculated from the above
equation, for ages of 62 and 93 days, which are the elapsed times between the
first and second $B$-band light-curve peaks and the mean epoch of the two
images.  It can be seen that the measured points fall into two families,
plotted for clarity as filled and open circles. The filled circles
satisfactorily follow the expected behavior if the corresponding echo rings are
illuminated by the first light-curve peak, while the open circles appear to
represent rings illuminated by the second peak. In strong confirmation of this
interpretation, we find that the filled circles all refer to the blue rings
seen clearly in Fig.~2b, while the open circles all correspond to the trailing,
red-edged rings.

Fig.~3 does show significant scatter, probably arising from departures of the
actual circumstellar structures from the simple adopted spherical geometry. For
plausible non-spherical geometries, measures made over several azimuths will
still average to a result close to the true distance. Thus, barring a very
unusual morphology, the figure implies that the distance to V838~Mon must
certainly be greater than 2~kpc, and probably significantly greater. Since the
predicted expansion rates approach an asymptotic limit with increasing
distance, we cannot make a stronger statement at present.  However, because the
relations for various distances diverge with increasing time, continued
high-resolution imaging will eventually yield a very precise distance.  

A second, independent geometrical method for determining the distance comes
from polarimetry\ref{8}{(Sparks 1994)}.  Since maximum linear polarization
occurs for $90^\circ$ scattering, there will be highly polarized light at a
linear radius $ct$, whose angular radius would thus yield the distance---but of
course only if there actually is dust at that radius and at the same distance
from us as the star.  Our most recent polarimetric \HST\/ image, obtained on
2002 September 2, shows very high polarization ($\sim$$50\%$), but only at the
inner rim of the cavity at the lower left (southeast) of the star. Thus, at
present, we can only set a lower limit to the distance, by equating the radius
at this rim, $\sim$$5\farcs0$, to $ct$, where $t$ is now the time since the
second light-curve peak.  This lower limit is $\sim$6~kpc. Future observations
will provide a definitive distance from this method as well, once the
light-echo paraboloid enlarges enough to allow us to illuminate close-in
material behind as well as in front of the star. At that time, we expect to see
maximum polarization in a ring\ref{8}{(Sparks 1994)}, rather than right at the
inner edge of the rim.

For a distance of at least 6~kpc (and interstellar reddening of
$E(B-V)\simeq0.8$\ref{16,23}{(Zwitter \& Munari 2002;  Kimeswenger et al.
2002)}), the absolute magnitude of V838~Mon at maximum was at least $M_V=-9.6$,
making it more luminous than a classical nova, and temporarily the brightest
star in the Milky Way.  The light curve, spectroscopic behavior, and high
luminosity of V838~Mon are reminiscent of ``M31 RV,'' a luminous red variable
star that appeared in the Andromeda galaxy in the late 1980's\ref{24}{(Mould et
al. 1990)}.  M31~RV reached an absolute visual magnitude at least as bright as
$M_V=-9.3$\ref{25}{(Bryan \& Royer 1992)}, but unfortunately was not well
studied before it declined below detectability.  V838~Mon and M31 RV (along
with, possibly, the Galactic object V4332 Sagittarii\ref{26}{[Martini et al.
1999]}) appear to represent a new type of outburst in which a star expands
rapidly to supergiant dimensions\ref{27}{(Munari et al. 2002b)}, but without
either the catastrophic envelope loss or the evolution to high temperature seen
in classical and symbiotic novae.  

In the case of V838~Mon the outburst appears to have occurred in a binary
system, which may suggest an analogy with a nova system in which a
thermonuclear explosion occurs on an accreting white-dwarf companion.  However,
a broad theoretical exploration\ref{28}{Prialnik \& Kovetz 1995} of the
parameter space for thermonuclear runaways on accreting white dwarfs predicts
no behavior like that of V838~Mon.  Evidently, V838~Mon has in the recent past
had several similar outbursts, which produced the dust envelope that is now
being illuminated. These previous outbursts appear to rule out one-time
catastrophic events such as a stellar collision or merger\ref{29}{Soker \&
Tylenda 2003}, or a common-envelope interaction\ref{30}{Iben \& Tutukov 1992},
as the energy source.  Now that V838~Mon appears to be returning to its
quiescent state, further observations of the system may shed light on its
properties and give insight into possible outburst mechanisms for this
mysterious and extraordinary object.



\clearpage

\noindent {\bf References:}


%
%
%

\def\AA#1{{\it Astron. Astrophys. \bf #1}}
\def\aasup#1{{Astron.~Astrophys. Suppl. \bf #1}}
\def\AJ#1{{\it Astron.~J. \bf #1}}
\def\ApJ#1{{\it Astrophys.~J. \bf#1}}
\def\MNRAS#1{{\it Monthly Notices Roy. Astron. Soc. \bf #1}}
\def\PASP#1{{\it Publ. Astron. Soc. Pacific \bf #1}}

\begingroup\parindent=0pt\frenchspacing
   \parskip=1pt plus 1pt minus 1pt\interlinepenalty=1000\pretolerance=10000
   \hyphenpenalty=10000\everypar={\hangindent=0.225in} 
\newcount\num
\def\enumerate{\num=0 }
\def\nextnum{\global\advance \num by 1 \number\num}
\def\nextitem{\leavevmode \noindent
   \hbox{\ifnum\num>8 \kern-0.43em\fi \nextnum.\kern0.60em}}

\enumerate

\nextitem Henden, A. et al. V838 Monocerotis. {\it IAU Circular} 7859 (2002).

\nextitem Munari, U. et al. The mysterious eruption of V838 Mon. \AA{389},
L51-L56 (2002).

\nextitem Schaefer, B.E. Light echoes: novae. \ApJ{327}, 347-349 (1988).

\nextitem Kapteyn, J.C. \"Uber die Bewegung der Nebel in der Umgebung von Nova 
Persei. {\it Astr. Nachrichten \bf 157}, 201-204 (1902).

\nextitem Couderc, P. Les aur\'eoles lumineuses des novae. {\it Ann.
d'Astrophys \bf 2}, 271-302 (1939).

\nextitem Russell, H.N., Dugan, R.S. \& Stewart, J.Q. {\it Astronomy: a
revision of Young's Manual of Astronomy} (Ginn \& Co., Boston, 1945), p. 786.

\nextitem Chevalier, R.A. The scattered-light echo of a supernova. \ApJ{308},
225-231 (1986).

\nextitem Sparks, W.B. A direct way to measure the distances of galaxies.
\ApJ{433}, 19-28 (1994).

\nextitem Ritchey, G.W. Nebulosity about Nova Persei. Recent photographs.
\ApJ{15}, 129-131 (1902).

\nextitem Casalegno, R.~et al.\ The emission nebula associated with V1974
Cygni: a unique object?  \AA{361}, 725-733 (2000).

\nextitem Sparks, W.B. et al. Evolution of the light echo of SN 1991T.
\ApJ{523}, 585-592 (1999).

\nextitem Bond, H.E., Gilmozzi, R., Meakes, M.G. \& Panagia, N. Discovery of
an inner light-echo ring around SN 1987A. \ApJ{354}, L49-L52 (1990).

\nextitem Brown, N.J. Peculiar Variable in Monoceros. {\it IAU Circular} 7785
(2002).

\nextitem Kato, T. http://vsnet.kusastro.kyoto-u.ac.jp/vsnet/index.html (2002).

\nextitem Goranskii, V.P. et al. Nova Monocerotis 2002 (V838 Mon) at early
outburst stages. {\it Astronomy Letters \bf 28}, 691-700 (2002).

\nextitem Kimeswenger, S., Lederle, C., Schmeja, S. \& Armsdorfer, B. The
peculiar variable V838 Mon. \MNRAS{336}, L43-L47 (2002).

\nextitem Osiwala, J.P. et al. The double outburst of the unique object V838
Mon. In {\it Symbiotic stars probing stellar  evolution}, eds. R.L.M. Corradi,
J. Mikolajewska \& T.J. Mahoney eds., (ASP  Conference Series, San Francisco,
2002), in press.

\nextitem Price, A. et al. Multicolor observations of V838 Mon. {\it IAU Inform.
Bull. Var. Stars} 5315, 1 (2002).

\nextitem Crause, L.A., Lawson, W.A., Kilkenny, D., van Wyk, F., Marang,
F. \& Jones, A.F. The post-outburst photometric behaviour of V838 Mon, \MNRAS{}
in press.

\nextitem Geballe, T.R., Smalley, B., Evans A. \& Rushton, M.T. V838
Monocerotis. {\it IAU Circular} 8016 (2002).

\nextitem Munari, U. \& Desidera, S. V838 Monocerotis. {\it IAU Circular} 8005
(2002).

\nextitem Wagner, R.M. \& Starrfield, S. V838 Monocerotis. {\it IAU Circular} 
7992 (2002).

\nextitem Zwitter, T. \& Munari, U. V838 Monocerotis. {\it IAU Circular} 7812
(2002).

\nextitem Mould, J. et al. A nova-like red variable in M31. \ApJ{353}, L35-L37
(1990).

\nextitem Bryan, J. \& Royer, R.E. Photometry of the unique luminous red
variable in M31. \PASP{104}, 179-181 (1992).

\nextitem Martini, P. et al. Nova Sagittarii 1994 1 (V4332 Sagittarii): The
discovery and evolution of an unusual luminous red variable star. \AJ{118},
1034-1042 (1999).

\nextitem Munari, U., Henden, A., Corradi, R.L.M. \& Zwitter, T. V838 Mon and
the new class of stars erupting into cool supergiants. In {\it Classical Nova 
Explosions}, eds. M.Hernanz \& J.Jose (American Institute of Physics, New York,
2002), p.~52-56. 

\nextitem Prialnik, D.~\& Kovetz, A.\  An extended grid of multicycle nova
evolution models. \ApJ{445}, 789-810 (1995).

\nextitem Soker, N.~\& Tylenda, R.\  Main-sequence stellar eruption model for
V838 Monocerotis. \ApJ{582}, L105-L108 (2003).

\nextitem Iben, I. \& Tutukov, A.V.\  Rare thermonuclear explosions in
short-period cataclysmic variables, with possible application to the nova-like
red variable in the Galaxy M31. \ApJ{389}, 369-374 (1992).



\endgroup


\acknowledgments

\noindent {\bf Acknowledgements} Based on observations with the NASA/ESA {\it
Hubble Space Telescope}, obtained at the Space Telescope Science Institute,
which is operated by AURA, Inc., under NASA contract NAS5-26555. We thank the
Space Telescope Science Institute for awarding Director's Discretionary
Observing Time for this project and for support through grant GO-9587. S.S. was
supported, in part, by NSF and NASA grants to Arizona State University.

\noindent{\bf Competing interests statement} The authors declare that they have
no competing financial interests.

\noindent{\bf Correspondence} and requests for materials should be addressed to
H.E.B. (e-mail: \discretionary{bond}{@stsci.edu}{bond@stsci.edu}).

\clearpage

\begin{deluxetable}{lcccc}
\tablewidth{0pt}
\tablecaption{\Hubble\/ ACS Wide-Field Channel Observing Log}
\tablehead{
\colhead{UT Date} & \colhead{HJD} & \colhead{Days since} & \colhead{Days since} &
\colhead{ACS filters} \\
\colhead{(2002)} & \colhead{ } & \colhead{Feb 6.4} & \colhead{Mar 10.0} &
\colhead{used}}
\startdata
Apr 30 & 2452394.7 & 82.8         & 51.2 & $B$+pol \\
May 20 & 2452415.5 & \llap{1}03.6 & 72.0 & $B$+pol, $V$+pol, $I$ \\
Sep 2  & 2452520.4 & \llap{2}08.5 & \llap{1}76.9 & $B$, $V$+pol, $I$ \\
Oct 28 & 2452576.0 & \llap{2}64.1 & \llap{2}32.5 & $B$, $V$, $I$ \\
\enddata
\tablenotetext{~}{The Advanced Camera for Surveys' F435W, F606W, and F814W
filters are denoted $B$, $V$, and $I$, respectively, and ``+pol'' denotes
that the indicated filter was combined in succession with 0-, 60-, and
120-degree polarizers. The time of maximum $B$ luminosity was near 2002
February 6.4, and the second $B$ peak was near March 10.0.
(The lack of \HST\/ observations between May and September
is due to solar avoidance.)
}
\end{deluxetable}

\clearpage


\begin{figure}
\begin{center}
%
%
\end{center}

\centerline{\bf FIGURE CAPTIONS}

\figcaption{Multicolor light curves of the outburst of V838~Mon.  Data have
been assembled by A.H. from observations by himself at the U.S. Naval
Observatory, Flagstaff, and from amateurs\ref{14}{(Kato 2002)} and other
professional observers\ref{2,15-19}{(Munari et al. 2002, Price et al. 2002,
Kimeswenger et al. 2002, Osiwala et al. 2002, Goranskii et al. 2002), Crause}.
Magnitudes in the standard Johnson blue ($B$), visual ($V$), and Kron-Cousins
near-infrared ($I$) bands have been converted to flux on a linear scale,
normalized to the peak at Heliocentric Julian Date 2452311.
Corresponding calendar dates are shown at the bottom. After an initial rapid
rise to 10th~magnitude and a subsequent gradual decline, V838~Mon abruptly rose
another 4 magnitudes in early February, reaching a maximum of visual magnitude
$V=6.75$ on 2002 February~6. After another decline, the star rose to a second
peak in early March. Once again there was a slow decline after this peak,
except in the near-infrared $I$ band, where yet a third peak was attained in
early April. In late April, V838~Mon dropped quickly back to its original
quiescent brightness in the blue ($B$) and $V$ bands; this precipitous drop in
brightness has been attributed to dust formation above the
photosphere\ref{19}{(Crause et al.\ 2003)}. The most recent observations show
some rebrightening in $I$ only. The progenitor object was recorded in archival
sky surveys\ref{2}{(Munari et al. 2002)} at visual magnitude~15.6, very similar
to its current post-outburst brightness.   }

\end{figure}
\clearpage


\begin{figure}
\begin{center}
\end{center}

\figcaption{\Hubble\/ Advanced Camera for Surveys images of the light echoes.
The apparently superluminal expansion of the echoes as light from the outburst
propagates outward into surrounding dust is shown dramatically.  Images were
taken on (a)~2002 April 30, (b)~May 20, (c)~September 2, and (d)~October 28.
Each frame is $83''\times 83''$; north is up and east to the left. Imaging on
April~30 was obtained only in the $B$ filter, but $B$, $V$, and $I$ were used
on the other three dates, allowing us to make full-color renditions. The time
evolution of the stellar outburst (Fig.~1) is dramatically reflected by
structures visible in the color images. In Fig.~2b, for example, note the
series of rings and filamentary structures, especially in the upper right
quadrant. Close examination shows that each set of rings has a sharp, blue
outer edge, a dip in intensity as one moves toward the star, and then a
rebrightening to a redder plateau.  Similar perfect replicas of the outburst
light curve are seen propagating outwards throughout all of the color images. }

\end{figure}
\clearpage


\begin{figure}
\begin{center}
\end{center}

\figcaption{Measured and predicted apparent angular expansion rates of rings
seen in the light echoes.  Measurements were done on the pair of $B$-band
images taken on 2002 Apr 30 and May 20.  A simple edge-finding algorithm was
used to mark the locations of the same rings on both images at several
different azimuths, and their angular expansion rates were calculated.  Filled
circles denote rings that we attribute to illumination from the first
light-curve peak, and open circles denote rings attributed to the second peak.
Measurement errors are only slightly larger than the size of the plotted
points. Units of expansion are milliarcseconds per day, as functions of angular
radius in arcseconds. Also shown are the predicted expansion rates for
light-echo rings seen 62 days (red dashed lines) and 93 days (blue solid lines)
after outburst, calculated from the equation given in the text. They are
functions of distance to the star, plotted for the range from 0.5 to 16 kpc,
and are apparently superluminal due to the parabolic geometry of the
illuminated surfaces. The measured angular expansion rates imply a strong lower
limit to the stellar distance of 2~kpc.}

\end{figure}
\clearpage


\begin{figure}
\begin{center}
%
%
\includegraphics[height=3.75in]{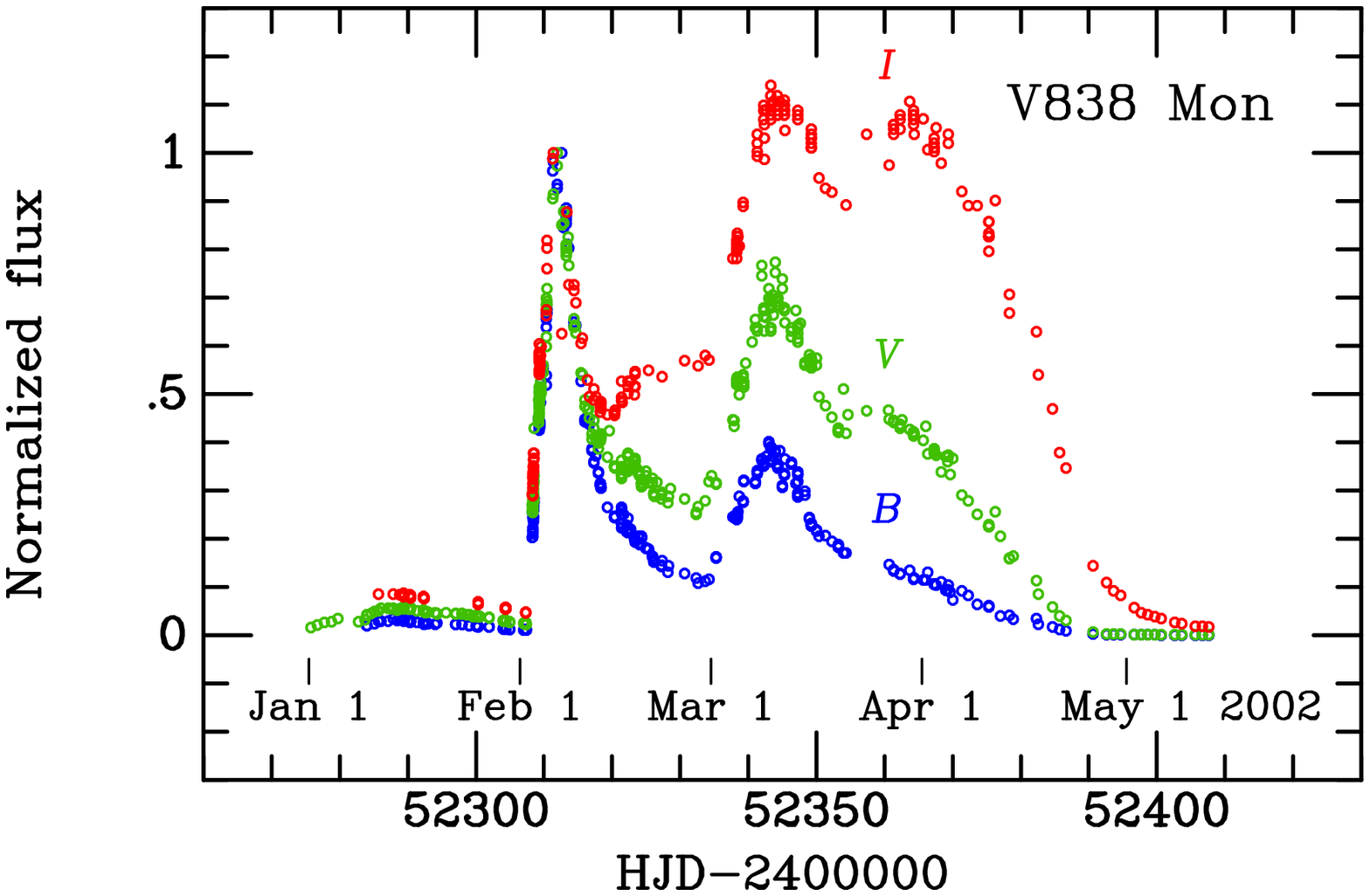}
\end{center}

\centerline{Figure 1}

\end{figure}
\clearpage


\begin{figure}
\begin{center}
\includegraphics[height=3.2in]{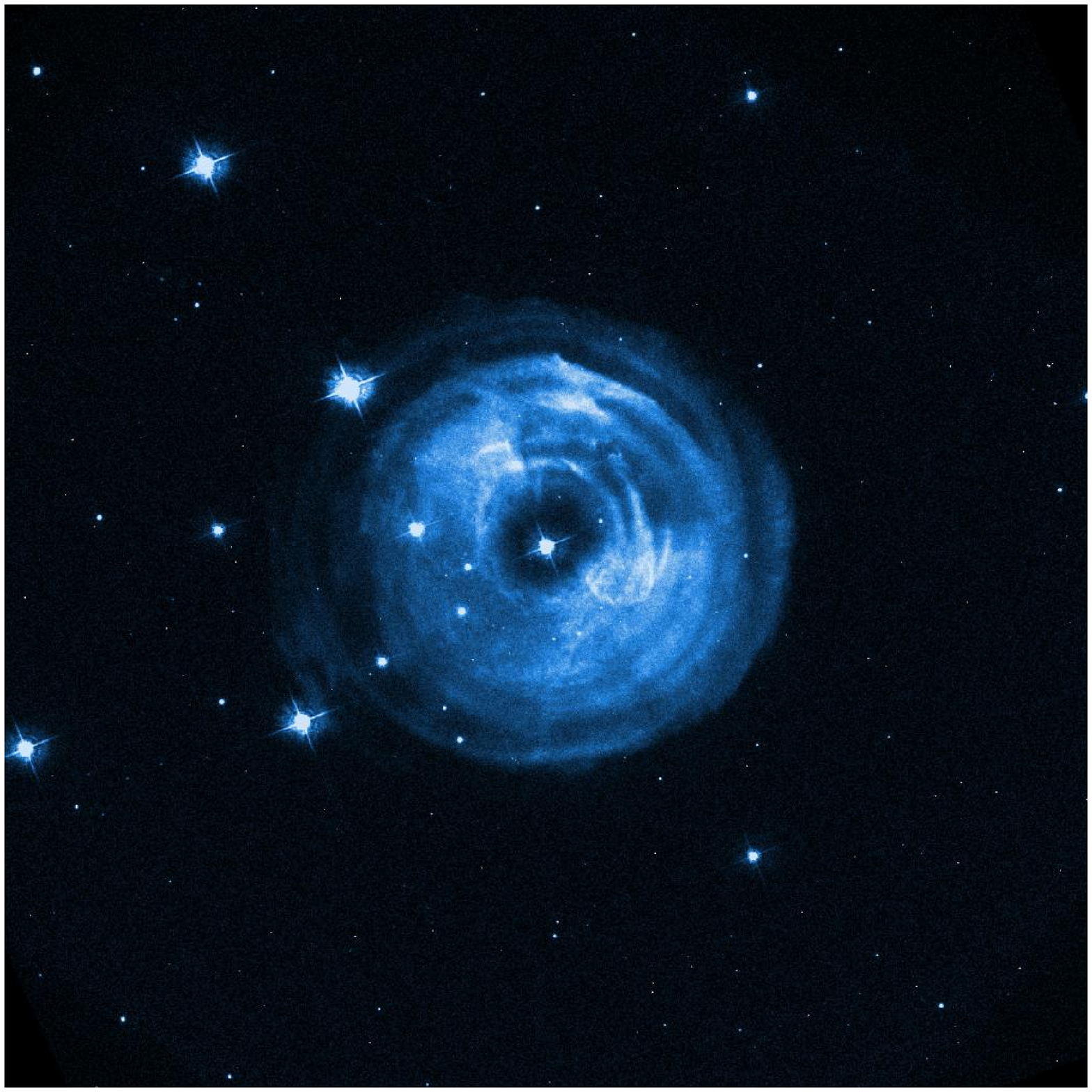}
\smallskip
\includegraphics[height=3.2in]{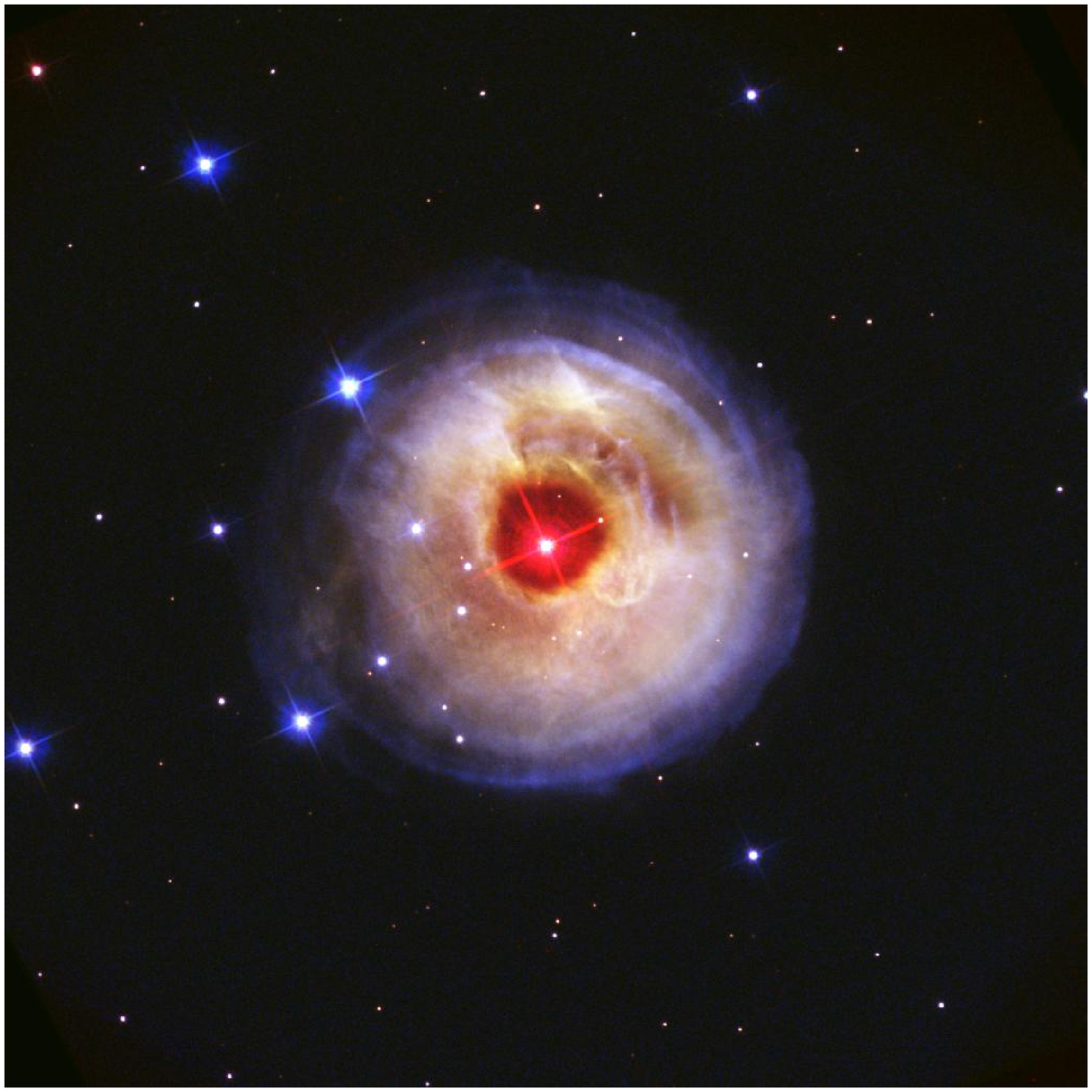}
\smallskip
\includegraphics[height=3.2in]{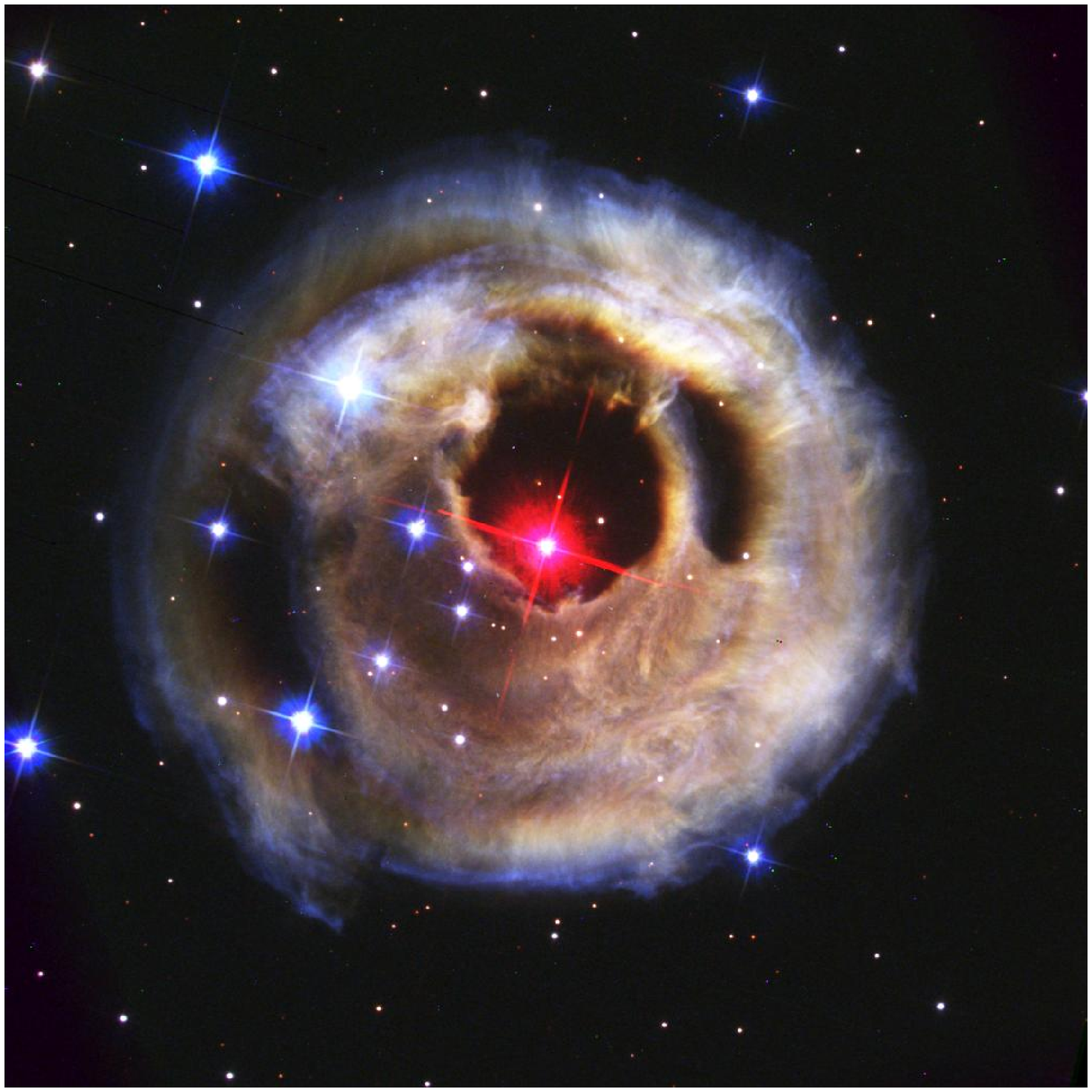}
\smallskip
\includegraphics[height=3.2in]{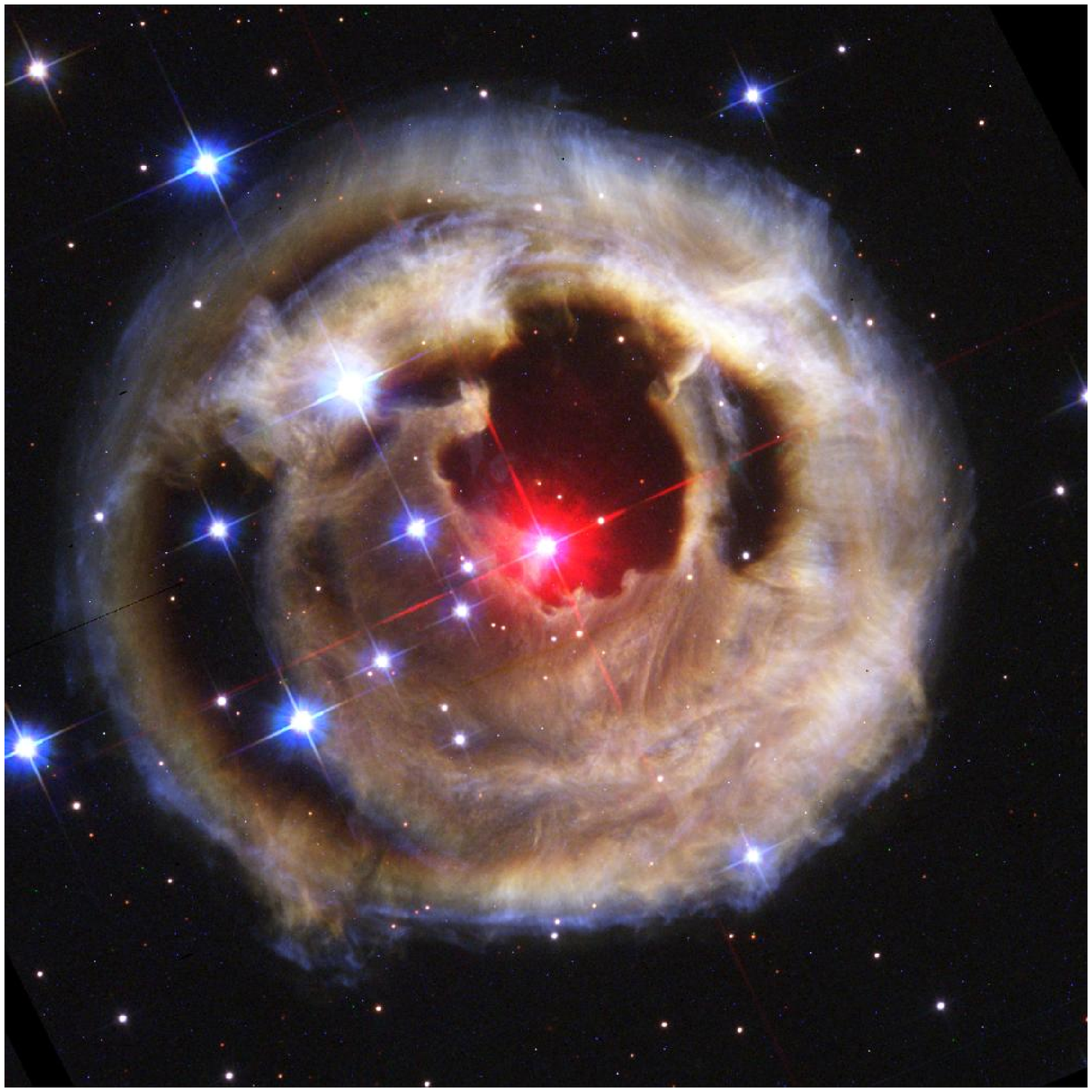}
\end{center}

\centerline{Figure 2}

\end{figure}
\clearpage


\begin{figure}
\begin{center}
\includegraphics[height=5in]{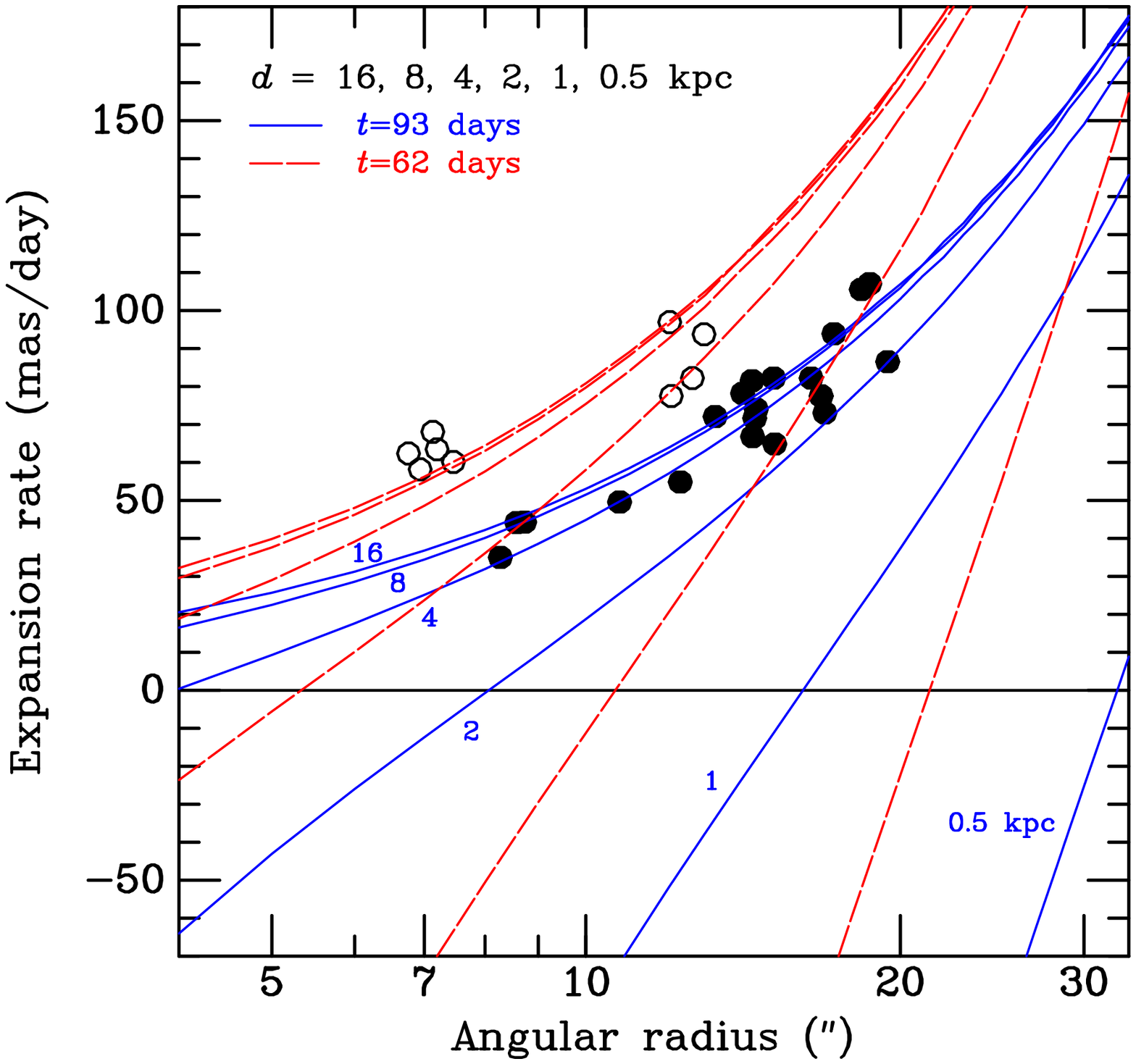}
\end{center}

\centerline{Figure 3}

\end{figure}
\clearpage

\end{document}